\documentclass{elsart1p}
\usepackage{graphicx}
\usepackage{amssymb}
\begin{document}
\begin{frontmatter}
%
%
%
%
%
\title{Inclusive production of $\pi^0$ in $pp$ collisions at 0.9 and
  7~TeV and perspectives for heavy-ion measurements with the ALICE
  calorimeters}
%
%
%
\author{Yuri Kharlov}
\author{for the ALICE collaboration}
\address{Institute for High Energy Physics, Protvino, 142281 Russia}
\begin{abstract}
  The inclusive spectrum of $\pi^0$ production has been measured in
  $pp$ collisions at $\sqrt{s}=900$~GeV and $7$~TeV with the ALICE
  experiment. The preliminary results of these measurements are
  presented, and perspectives for $\pi^0$ measurements with heavy ions
  are discussed.
\end{abstract}
\end{frontmatter}

\section{Introduction}
\label{sec:Introduction}

Hadron production measurements in proton-proton collisions at the LHC
energies opens up a new kinematic regime for testing and
validating the predictive power of quantum chromodynamics, and to
impose new constraints on models and their parameters. Quantitative
description of hard processes is provided by perturbative QCD
(pQCD). However, a significant fraction of hadrons are produced in
$pp$ collisions at high energies via soft parton interactions, and
thus they cannot be well described within the framework of pQCD. Many
advanced event generators have to appeal to phemonenological models,
along with the pQCD calculations, in order to describe hadron
production adequately. Evidently, such phenomenological models are
tuned to available experimental data, and have been validated using
data delivered by lower-energy colliders, like RHIC, SPS and
Tevatron. Extrapolation of these models to LHC energies cannot be
valid {\it a priory}, because the increase in the collision energy is
very large. Even the validity of the pQCD predictions cannot be
guaranteed at the LHC, since the parton density functions (PDF) are
not well determined at such a high energy.

Hadron spectra measured in heavy ion collisions shed light onto the
parton energy loss in hot quark-gluon matter, via comparison with the
spectra measured in $pp$ collisions. Suppression of the hadron yield,
defined as the ratio of the hadron production spectra in central heavy
ion and $pp$ collisions, normalized per nucleon-nucleon collision, is
referred to as the nuclear modification factor $R_{AA}$. It was
measured in Au-Au collisions at $\sqrt{s_{_{NN}}}=200$~GeV by
PHENIX~\cite{PHENIX} and STAR~\cite{STAR} at RHIC. The LHC brings the
Pb-Pb collisions to almost 10 times higher energy,
$\sqrt{s_{NN}}=2.76$~TeV, and thus the measurements of $R_{AA}$
becomes important for understanding the properties of the quark matter
produced in these high-energy nuclear collisions. Suppression of
charged particle production at large $p_T$ in central Pb-Pb collisions
at $\sqrt{s_{_{NN}}}=2.76$~TeV has been already observed by
ALICE~\cite{ALICE_RAA}.

The ALICE experiment \cite{ALICE} performs measurements of the neutral
pion production in $pp$ collisions at the collision energies
$\sqrt{s}=7$~TeV and 900~GeV, and in Pb-Pb collisions at
$\sqrt{s_{_{NN}}}=2.76$~TeV at mid-rapidity in a wide range of the
tranvserse momenta $p_T$. Conventionally, the $\pi^0$ meson is detected
via its two-photon decay in the electromagnetic calorimeters, PHOS and
EMCAL. The PHOS detector \cite{PHOS} covers the acceptance of
$260^\circ<\varphi<320^\circ$ in azimuth angle and $|\eta|<0.13$ in
pseudorapidity. The EMCAL \cite{EMCAL} acceptance is
$80^\circ<\varphi<120^\circ$ and $|\eta|<0.7$. The decay
$\pi^0\to\gamma\gamma$ was also measured by ALICE via identifying the
conversion photons produced in the material of the ALICE inner
tracking system, $\gamma\to e^+e^-$, described elsewhere
\cite{AamodtHQ2010,KochHP2010}.

\section{Analysis}

The proton-proton collision data used for the measurements of
the $\pi^0$ spectrum were collected by the ALICE detector in 2010 with
the minimum bias trigger \cite{ALICE-2009-01}. This trigger required a
hit in the Silicon Pixel Detector (SPD) or in either one of the two
scintillator hodoscopes V0A and V0C surrounding the interaction point
at large rapidities. The integrated luminosities of the event samples
are $\int{\mathcal L}dT = 5.5~\mbox{nb}^{-1}$ at $\sqrt{s}=7$~TeV and
$\int{\mathcal L}dT = 0.14~\mbox{nb}^{-1}$ at $\sqrt{s}=900$~GeV.

\begin{figure}[ht]
  \hfil
  \includegraphics[width=0.38\hsize,bb=0 40 567 530]{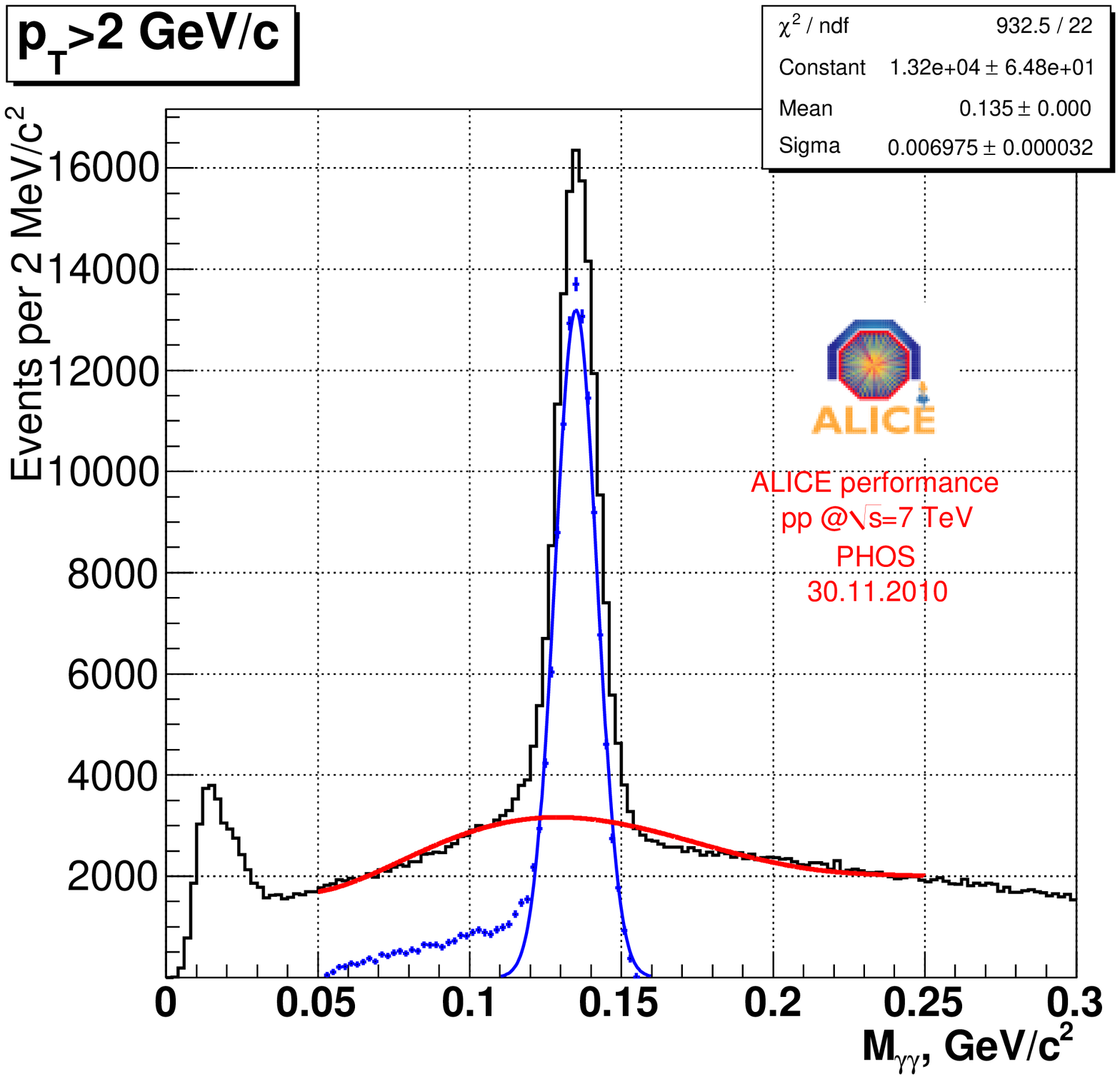}
  \hfil
  \includegraphics[width=0.38\hsize,bb=0 40 567 530]{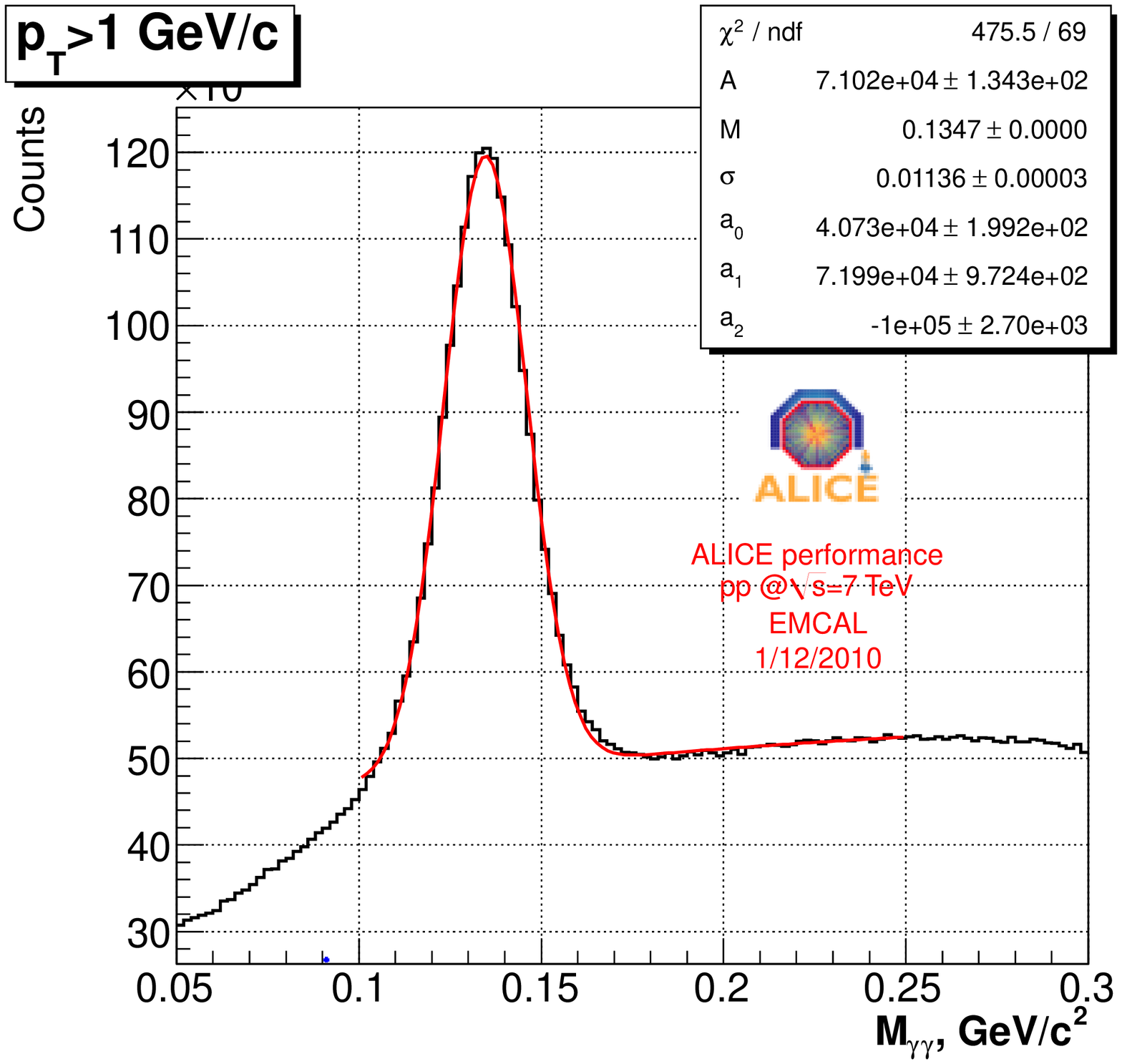}
  \hfil
  \caption{Invariant mass spectra of cluster pairs measured in PHOS
    and EMCAL.}
  \label{fig:InvMass}
\end{figure}
Reconstruction of the $\pi^0$ mesons in the ALICE calorimeters, PHOS
and EMCAL, was performed by invariant mass analysis. To minimize a
possible bias by the photon identification, rather loose cuts on the
photon candidates were imposed. To suppress a major part of hadronic
background, the lower cut on the cluster energy was set to a value
just above the minimum ionizing energy, $E>0.3$~GeV. An additional cut
on the number of cells in a cluster was set in PHOS --- all clusters
containing at least 3 cells were considered as candidates for
photons. Due to the low occupancy of both calorimeters by the secondary
particles in $pp$ collisions, the background under the $\pi^0$ peak is
not very large and allows easily to extract the number of
$\pi^0$'s. Examples of the invariant mass distributions in PHOS and
EMCAL are shown in Fig.\ref{fig:InvMass}.  The number of reconstructed
$\pi^0$'s was found in each $p_T$ bin from the invariant mass
distributions by fitting and extracting the number of events under the
$\pi^0$ peak. The raw spectrum obtained was corrected for the
reconstruction efficiency calculated in Monte Carlo simulations tuned
to reproduce the real-data characteristics.

\section{Results and discussion}
\vspace*{-5pt}

\begin{figure}[ht]
  \hfil
  \includegraphics[width=0.38\hsize,bb=0 50 567 540]{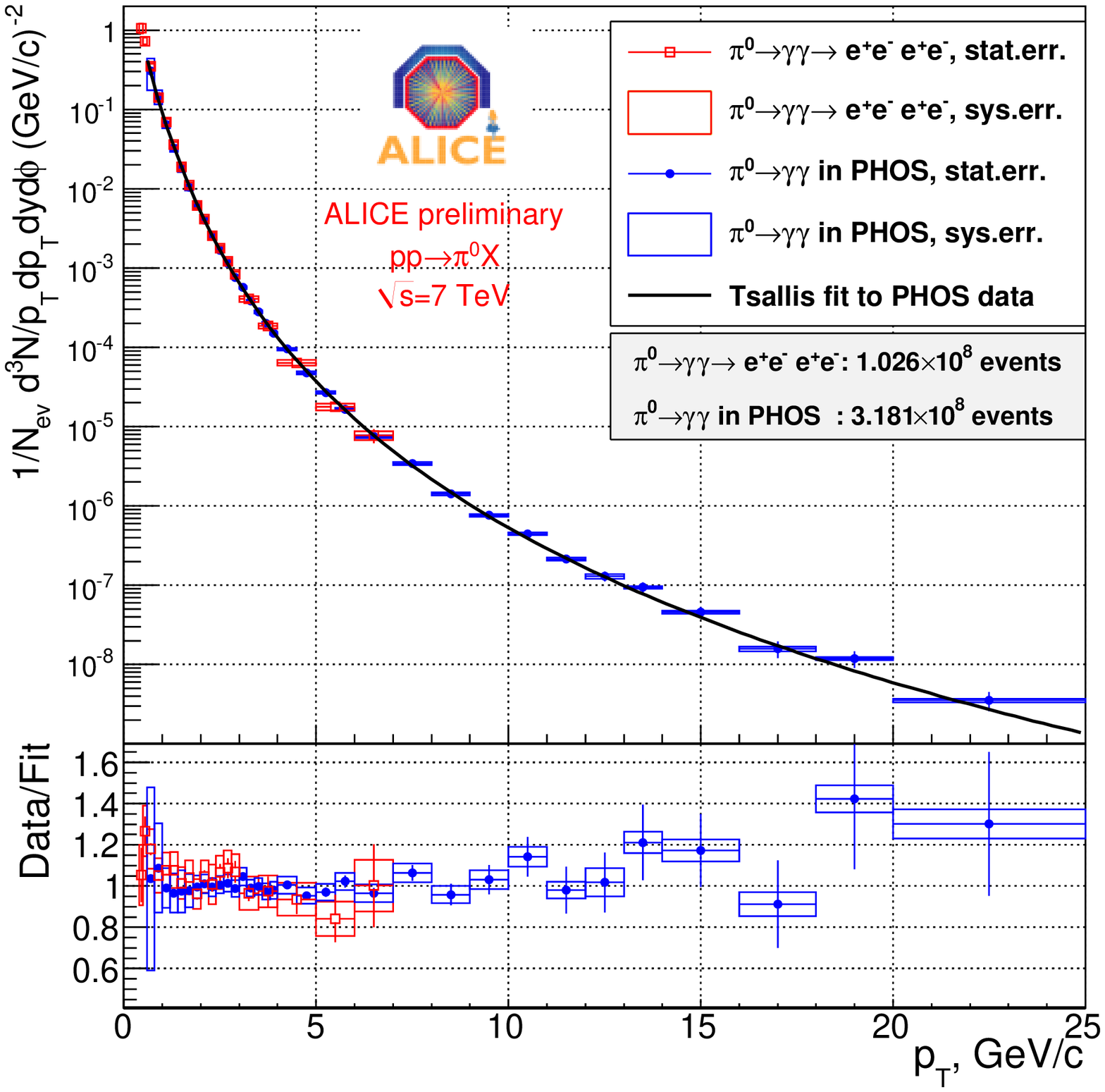}
  \hfil
  \includegraphics[width=0.38\hsize,bb=0 50 567 540]{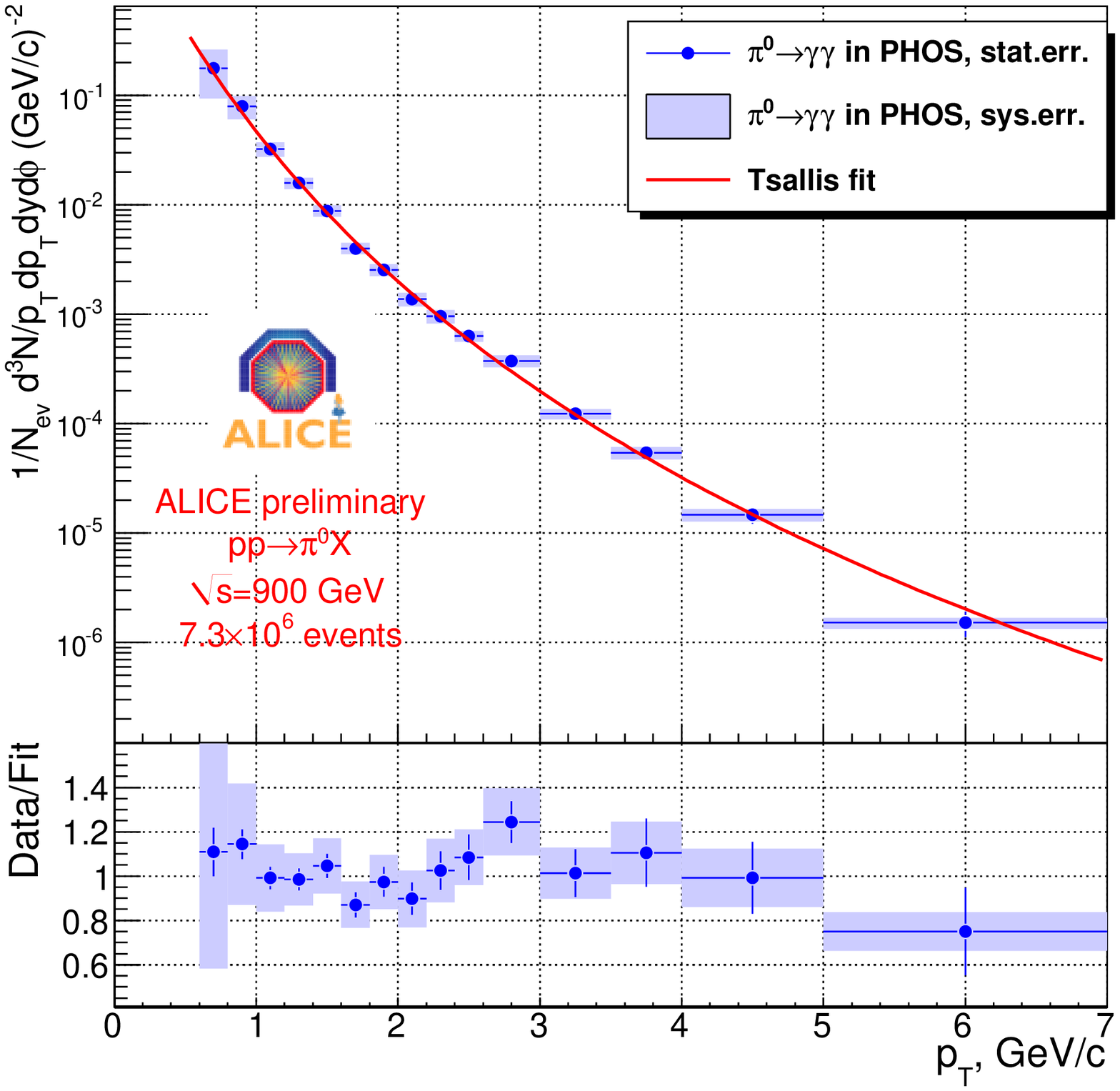}
  \hfil
  \caption{Normalized invariant production yield of $\pi^0$ mesons in
    $pp$ collisions at $\sqrt{s}=7$~TeV (left) and 900~GeV (right).}
  \label{fig:YieldNormalized}
\end{figure}

\begin{figure}[ht]
  \hfil
  \includegraphics[width=0.38\hsize,bb=0 50 567 540]{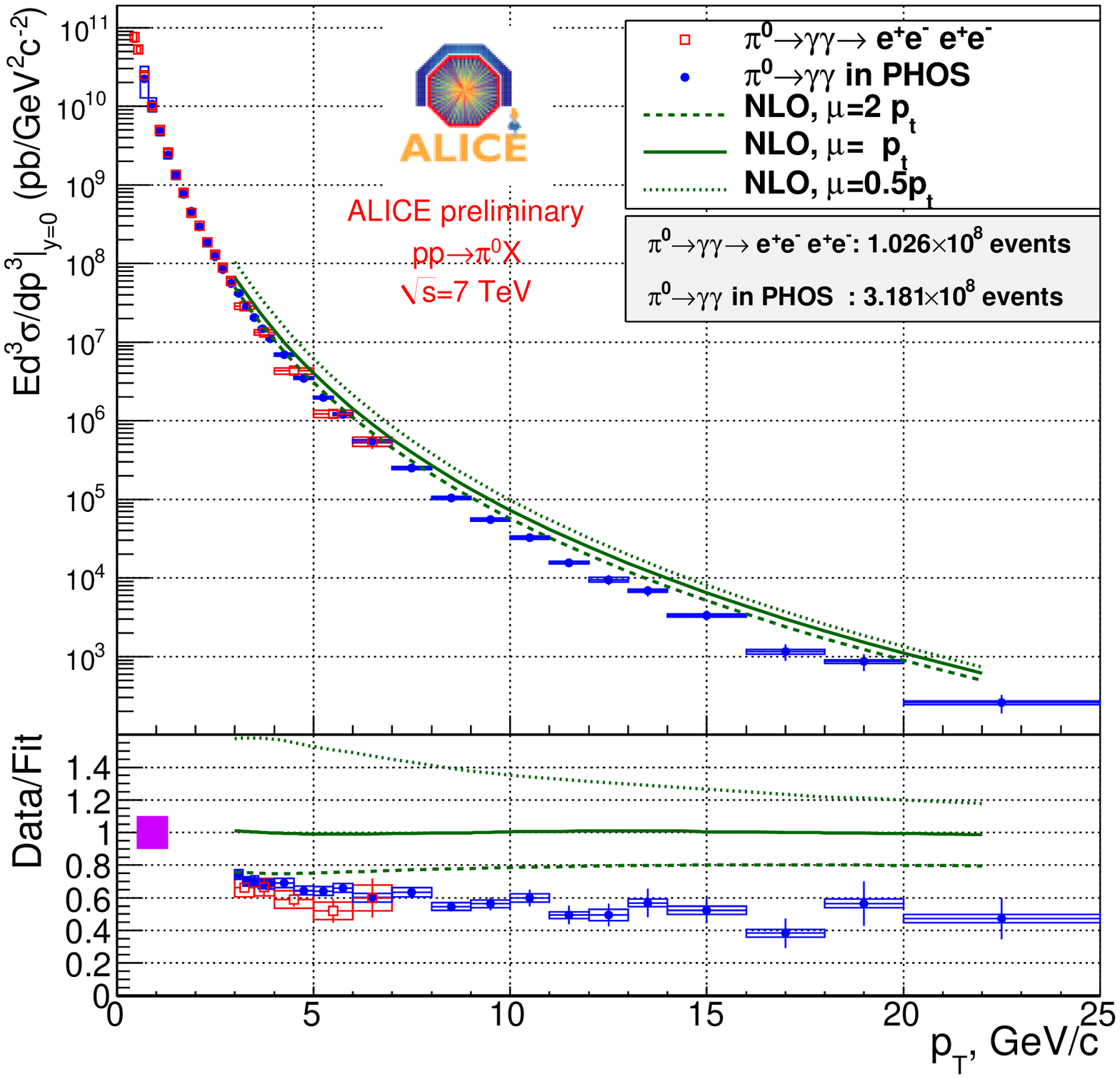}
  \hfil
  \includegraphics[width=0.38\hsize,bb=0 50 567 540]{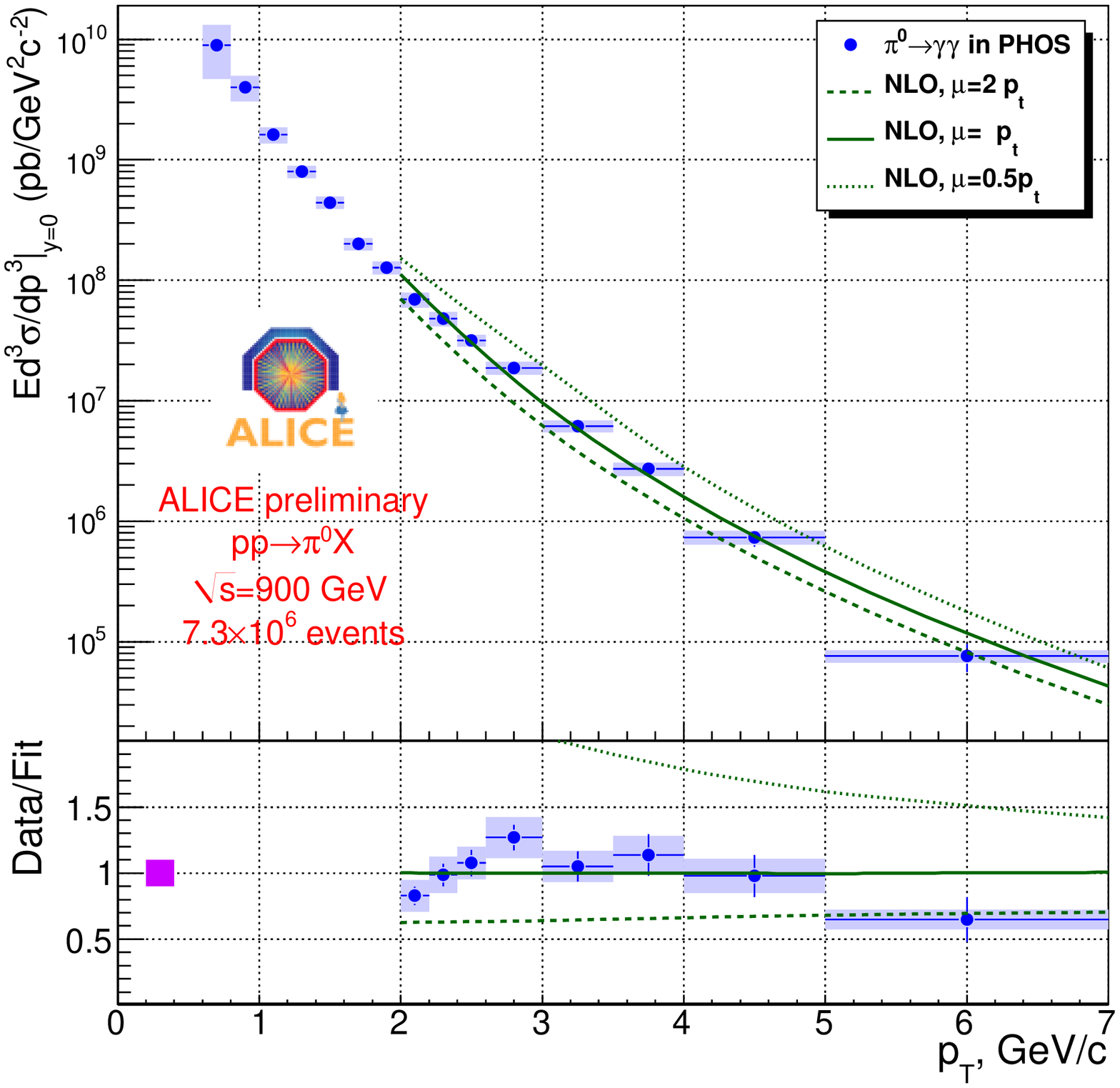}
  \hfil
  \caption{Differential invariant cross section of $\pi^0$ production
    in $pp$ collisions at $\sqrt{s}=7$~TeV (left) and 900~GeV
    (right).}
  \label{fig:CrossSection}
\end{figure}

Data collected with the PHOS detector in the $pp$ run allowed to
measure the $\pi^0$ spectra in the $p_T$ range from 0.6 to 25~GeV/$c$
at the center-mass energy $\sqrt{s}=7$~TeV and in the $p_T$ range from
0.6 to 7~GeV/$c$ at $\sqrt{s}=900$~GeV. The invariant $\pi^0$
production yields normalized per $pp$ minimum bias collision are shown
for both collision energies in Fig.\ref{fig:YieldNormalized}.
Besides the PHOS measurements, these plots show the results of the
measurements in the central tracking system via photon conversion. The
PHOS points were fitted by the Tsallis function $d^3N/p_tdp_{t}dyd\phi
= C[1 + (m_{T}-m)/nT]^{-n}$ and the ratio of the data points to the
fitting function, shown at the bottom of the spectra, illustrates the
stability of the measured points.  These normalized spectra were
converted to the invariant differential cross section of the $\pi^0$
production $Ed^3\sigma/dp^3|_{y=0}$ with the assumption of the
absolute cross section of $pp$ collisions. Within conservative
uncertainty estimation, the $pp$ cross section was taken as
$\sigma_{pp}=67\pm 10$~mb at $\sqrt{s}=7$~TeV and $\sigma_{pp}=50\pm
10$~mb at $\sqrt{s}=900$~GeV. The production cross sections obtained 
are shown in Fig.\ref{fig:CrossSection}.
Next-to-Leading Order pQCD calculations with the parton density
function CTEQ5M, fragmentation function KKP and different QCD scales
$\mu$ \cite{INCNLO} have been compared with the data. The ratio of the
measured cross section to the NLO prediction is shown in
the bottom panels. The uncertainty in the $pp$ cross section $\pm
10$~mb is represented by the pink box. At the collision energy of
900~GeV the NLO calculations at $\mu=p_T$ describe well the measured
data, while at $\sqrt{s}=7$~TeV the higher QCD scale ($\mu>2p_T$) is
required to reproduce better the data, although the discrepancy is
still significant.

Data collected by the ALICE calorimeters in Pb-Pb collisions at
$\sqrt{s_{NN}}=2.76$~TeV are sufficient to measure the $\pi^0$ spectrum
at $1<p_T<15$~GeV/$c$. More sophisticated analysis is ongoing now
involving advanced methods of background subtraction and of
reconstruction efficiency evaluation in a high-multiplicity
environment. The nuclear modification factor $R_{AA}$ of the $\pi^0$
spectrum will be a complimentary measurement to the charged particle
suppression \cite{ALICE_RAA} which will lead to better understanding
of transport propereties of hot QCD matter.

\section*{Conclusion}
\vspace*{-5pt}

The ALICE experiment at the LHC has measured the production spectrum
of neutral pions in proton-proton collisions at the energies
$\sqrt{s}=7$~TeV and 900~GeV, using two independent methods. The
photons from the $\pi^0$ were detected by the calorimeters, as well as
via photon conversion identified in the central tracking
system. Deploying these techniques provided a cross check and allowed
to reduce systematic uncertainties in the overlapping $p_T$ region and
to extend the joint spectrum to a wide $p_T$ range. The production
yield was measured at mid-rapidity at $0.4<p_T<25$~GeV/$c$ at
$\sqrt{s}=7$~TeV and at $0.6<p_T<7$~GeV/$c$ at
$\sqrt{s}=900$~GeV. These measurements allow a test of pQCD-based
calculations and provide reference data to measure the nuclear
modification factor $R_{AA}$ of the $\pi^0$ production in heavy ion
collisions at the LHC.

This work was partially supported by the grants RFBR 08-02-91021 and
10-02-91052. 



\begin{thebibliography}{00}
\vspace*{-5pt}
\bibitem{PHENIX} 
  A.Adare et al., PHENIX Collaboration. Phys.Rev.Lett.\ 101 (2008) 232301. 
\bibitem{STAR}
  J. Adams et al., STAR Collaboration, Phys. Rev. Lett.\ 91 (2003) 172302.
\bibitem{ALICE_RAA}
  K. Aamodt et al. ALICE collaboration. Phys. Lett. B 696 (2011) 30--39.
\bibitem{ALICE} 
  K. Aamodt et al. ALICE collaboration. JINST. 2008, 3, S0800.
\bibitem{PHOS}
  ALICE collaboration. Photon Spectrometer PHOS, Technical Design
  Report. CERN/LHCC 99-4, 5 March 1999.
\bibitem{EMCAL}
  ALICE collaboration. The electromagnetic calorimeter, Addendum to
  the Technical Proposal. CERN/LHCC 2006-014, 14 April 2006.
\bibitem{AamodtHQ2010}
  K.Aamodt et al. In: 4th Hot Quarks Workshop, La Londe Les Maures, 21
  June 2010. 
\bibitem{KochHP2010} 
  K.Koch et al. In: Conference Hard Probes 2010, Eilat, 6--10 October
  2010. To appear in Nucl.Phys.A.
\bibitem{ALICE-2009-01}
  K. Aamodt et al. ALICE collaboration. Phys.Lett.\ B 693 (2010) 53--68.
\bibitem{INCNLO}
  P. Aurenche, et al., Eur. Phys. J. C 13,347 (2000)
\end{thebibliography}
\end{document}